\documentclass[twocolumn]{revtex4}

\usepackage{amsmath,amssymb,amsthm}
\usepackage{graphicx}
\usepackage{epsfig}

\begin{document}
\title{Entangled gene regulatory networks with cooperative expression endow robust adaptive responses to unforeseen environmental changes} 

\author{Masayo Inoue\textsuperscript{1}, Kunihiko Kaneko\textsuperscript{2,3}} 

\affiliation{$^1$ School of Interdisciplinary Mathematical Sciences, Meiji University, Tokyo, Japan \\
$^2$ Department of Basic Science, Graduate School of Arts and Sciences, University of Tokyo, Tokyo, Japan \\
$^3$ Center for Complex Systems Biology, Universal Biology Institute, University of Tokyo, Tokyo, Japan}

\begin{abstract} 
Living organisms must respond to environmental changes. Generally, accurate and rapid responses are provided by simple, unidirectional networks that connect inputs with outputs. Besides accuracy and speed, however, biological responses should also be robust to environmental or intracellular noise and mutations. Furthermore, cells must also respond to unforeseen environmental changes that have not previously been experienced, to avoid extinction prior to the evolutionary rewiring of their networks, which takes numerous generations. To address the question how cells can make robust adaptation even to unforeseen challenges, we have investigated gene regulatory networks that mutually activate or inhibit, and have demonstrated that complex entangled networks can make appropriate input-output relationships that satisfy such adaptive responses. Such entangled networks function when the expression of each gene shows sloppy and unreliable responses with low Hill coefficient reactions. To compensate for such sloppiness, several detours in the regulatory network exist. By taking advantage of the averaging over such detours, the network shows a higher robustness to environmental and intracellular noise as well as to mutations in the network, when compared to simple unidirectional circuits. Furthermore, it is demonstrated that the appropriate response to unforeseen environmental changes, allowing for functional outputs, is achieved as many genes exhibit similar dynamic expression responses, irrespective of inputs including unforeseen inputs. The similarity of the responses is statistically confirmed by applying dynamic time warping and dynamic mode decomposition methods. As complex entangled networks are commonly observed in the data in gene regulatory networks whereas global gene expression responses are measured in transcriptome analysis in microbial experiments, the present results give an answer how cells make adaptive responses and also provide a novel design principle for cellular networks.
\end{abstract}

\maketitle

\section*{Author summary}
Recent experimental advances have demonstrated that cells often have appropriate, robust responses to environmental changes, including those that have not previously been experienced. It is known that accurate and rapid responses can be achieved by simple unidirectional networks that connect straightforwardly input and outputs. However, such responses were not robust to perturbations. Here we have made numerical evolution of gene regulatory networks with mutual activation and inhibitions, and uncovered that complex entangled networks including many feedforward and feedback paths can make robust input-output responses, when each gene expression is not accurate. Remarkably, they make appropriate responses even to unforeseen environmental changes, as are supported by global, correlated responses across genes that are similar for all input signals. The results explain why cells adopt complex gene regulatory networks and exhibit global expression changes, even though they may not be advantageous in terms of their energy cost or response speed. The present results are consistent with the recent experiments on microbial gene expression changes and network analyses. This investigation provides insights into how cells survive fluctuating and unforeseen unpredictable environmental changes, and gives a universal conceptual framework to go beyond the standard picture based on a combination of network motifs.

\section*{Introduction}

Living organisms, generally respond appropriately to environmental changes by adapting their internal states to the new conditions. 
In cells, chemical compositions, in particular gene expression levels, are adjusted to adapt to novel environments. 
Information on the environmental change is transmitted via a signal transduction network to regulate gene expression patterns. 

For a cell to make an appropriate response rapidly and accurately, it will be relevant to transfer the input signal to the output expression pattern unidirectionally through a sequence of chemical responses. Indeed, such unidirectional circuits are ubiquitously observed as direct connections or feed-forward networks from the input in the signal transduction networks. How these unidirectional networks achieve appropriate input-output relationships has previously been extensively investigated~\cite{Mangan:2003aa,Alon:2006aa,Karlebach:2008aa}. 

Real cellular networks, however, are often much more complex. They include many components and are entangled. Sometimes such unidirectional circuits may be extracted as a part of the complex network, but it is not clear whether they have the same function as has been determined in isolation~\cite{Ma:2004aa,Jimenez:2017aa}. They generally do not function independently, as the networks are entangled. 
Why cells have adopted such complex networks, however, is not completely understood.

There are further requirements for cellular responses beyond being quick and accurate. For instance, they should be robust to perturbations~\cite{Kaneko:2006ac,Ciliberti:2007aa,Kaneko:2007aa,Wagner:2007ab,Nagata:2020aa}, as the environmental conditions surrounding living organisms normally fluctuate~\cite{Elowitz:2002aa,Furusawa:2005aa,Hashimoto:2016aa}. 
Moreover, each process controlling the cellular concentrations of chemicals is generally stochastic, as the number of molecules involved is sometimes not very large~\cite{Swain:2002aa,Eldar:2010aa,Taniguchi:2010aa}. 
In addition to the robustness to fluctuations and noise, there is another postulate, an `adaptive response to an unforeseen challenge (AUC)', as put forward by Braun~\cite{Braun:2015aa}:  
Cells must manage to respond to and survive under unforeseen environmental changes that are not expected by the prescribed input-output responses. 
In particular, even when cells experience a novel environment for which appropriate output responses are not yet prepared, they have to survive to some degree to avoid extinction, as it takes many generations for the appropriate evolutionary changes to occur. 
Responses to unforeseen challenges are thus required without rewiring the gene regulatory network (GRN) via genetic evolution~\cite{Stolovicki:2006aa,David:2010aa,David:2013aa}. 

Direct unidirectional networks optimized for given input-output relationships are not expected to be robust. Further, it will be difficult to achieve AUCs with them. 
If this is the case, and if a complex network can have advantages over unidirectional networks, with regard to robustness and AUCs, this may explain why biological networks are so complex and entangled. However, it is not yet evident if a complex network with many degrees of freedom acquires such robustness and AUCs, as they do not always have functional robustness. It is thus necessary to understand which types of complex networks can have a higher capacity for robustness and AUCs than simple unidirectional networks and to uncover design principles for such networks.

Previously we have evolved GRNs to achieve appropriate input-output relationships. In addition to simple direct or feed-forward networks, we have uncovered another type of entangled network consisting of many components, that can generate an appropriate input-output relationship as a cooperative response of many genes~\cite{Inoue:2018aa}. In particular, this cooperative response emerges when each gene expression is sloppy, that is, with a low Hill coefficient as observed in a real GRN~\cite{Becskei:2005aa,Rosenfeld:2005aa,Dekel:2005aa,Kim:2008aa}. 
Here, we explore whether this class of networks achieves higher capacity in noise robustness and AUCs than traditional direct or feed-forward networks.

\section*{Results}

\begin{figure}[!h]
\centering
\includegraphics[width=8cm]{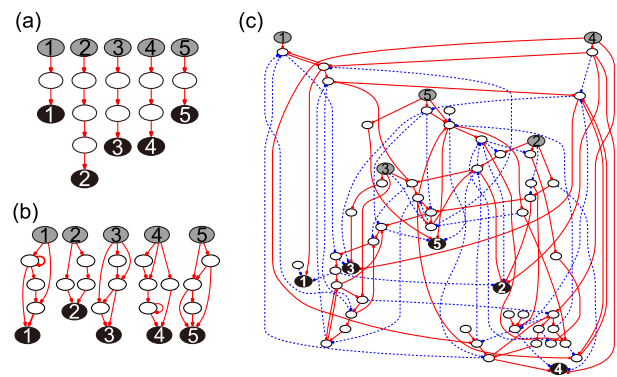}
\caption{{\bf Typical examples of the three types of core network structures.}
The core structures were obtained by removing the dispensable paths from the evolved fittest structure for each type. Gray and black circles show the input and output genes, respectively, and the numbers indicate their correspondences. Red bold arrows represent excitatory regulations, and blue dotted arrows denote inhibitory regulations. (a) Direct type with $y_T = 0.5$ and $\beta = 10^{1.75}$, (b) FF-network type with $y_T = 1.0$ and $\beta = 10^{1.5}$, and (c) Cooperative type with $y_T = 0.5$, and $\beta = \sqrt{10}$.}
\label{fig_core}
\end{figure}

We adopted a simplified GRN model~\cite{GLASS:1973aa,MJOLSNESS:1991aa,Salazar-Ciudad:2001aa,Kaneko:2007aa,Furusawa:2008aa,Inoue:2013aa,Inoue:2018aa} and investigated the networks evolved so that, for each given input, one prescribed output gene (target) is expressed, by defining the fitness of the expression patterns of the given output genes (Eq~(\ref{eq_fitness})). 
Each input stimulated one of the input genes and was transmitted to the output genes through the middle-layer genes (ML-genes). As described in~\cite{Inoue:2018aa}, three types of networks evolved depending on the sensitivity (corresponding to the Hill coefficient) of the gene expression dynamics: 
direct connections (Direct type) and feed-forward networks having side paths (FF-network type) that showed rapid and accurate responses evolved when each gene's response was sensitive (large $\beta$), whereas complex entangled networks with cooperative responses of many genes (Cooperative type) evolved when each response was sloppy (small $\beta$). 
The three types showed different characteristics in their network structures, and these were most discernible when their core structures were compared. 
The core structure was obtained by removing paths successively, one-by-one, as long as the corresponding output genes response for each input was preserved (e.g., maintaining fitness $>0.8$) (Fig~\ref{fig_core}). 
The Direct and FF-network types adopted unidirectional circuits from the input to the output genes. The Cooperative network type, however, involved many genes and its network was entangled. In this case, with the input, a few genes were locally excited, whereas global inhibition by many other genes followed.
The characteristic structure of local excitation and global inhibition (LEGI~\cite{Levchenko:2002aa}) was common in this Cooperative type. 
Here, it is of note that the evolutionary simulations were carried out to preserve the total path number of the networks, and the redundant paths out of the core structure remained for all types. 

\subsection*{Noise robustness}

\begin{figure*}[htb]
\centering
\includegraphics[width=13.2 cm]{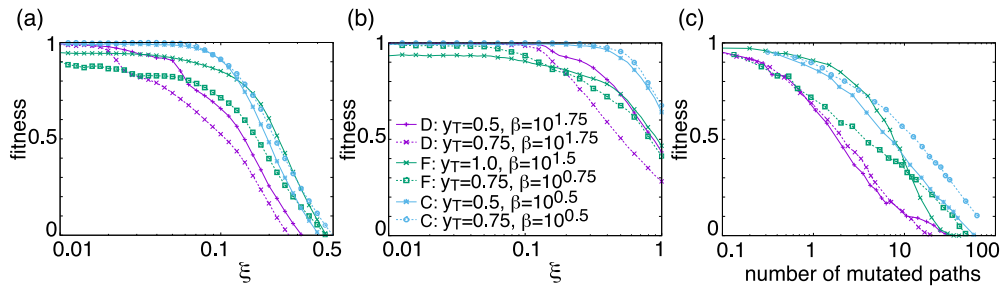} 
\caption{{\bf Noise robustness.}
(a, b) Dependence of fitness (ordinate) on noise strength $\xi$ (abscissa). The average fitness values for each noise level are plotted using 5 different networks that evolved without noise for each parameter set. Noise was added to all ML-genes (a) or to an input gene stimulated by an external input (b). (c) Dependence of the fitness (ordinate) on mutation, i.e., the expected number of removed paths from the core structures (abscissa). Results of the Direct type (D: purple $+$ and $\times$), the FF-network type (F: green $\times$ and $\square$), and the Cooperative type (C: blue $\ast$ and $\circ$) are shown.}
\label{fig_mut}
\end{figure*}

Robustness to noise was investigated first. A white Gaussian noise term with amplitude $\xi$ was added to the time evolution of each expression level (Eq~(\ref{eq_dynamics})) for the ML-genes (Fig~\ref{fig_mut} (a)) or stimulated input genes (Fig~\ref{fig_mut} (b)), and the corresponding Langevin equations were simulated. 
When the noise term was added to the ML-genes, the Direct type was the most fragile and the fitness was drastically reduced by the noise (Fig~\ref{fig_mut} (a)). The FF-network and the Cooperative types showed noise robustness, as their signal transduction routes were multitracked and noise could be canceled out along the pathways. 
Furthermore, the Cooperative type was more robust to smaller amounts of noise, with $\xi < 0.1$. This was because the entangled network enabled the averaging out of noise, not only on each gene at the endpoint of multitrack paths but also on every gene in the network. It is of note that the noise canceling effect was independent of how many genes affected each output gene. 
An output gene had interaction paths with $6.1$ genes in the Direct type, $13.0$ genes in the FF-network type, and $11.0$ genes in the Cooperative type, on average. 
In addition, the total number of paths in a network was set to be the same for all types. Hence, there was no correlation between the number of paths of interacting genes and noise robustness. 

We also studied robustness against noise to an input gene (Fig~\ref{fig_mut} (b)). 
Again, the Direct type was largely influenced by the input noise as the noise effect was transmitted directly via the straight paths. The FF-network type was less robust than the Direct type. This is because its feed-forward network amplifies the input noise as well as the external signal. The Cooperative type had higher robustness to the input noise. Again, the noise could be reduced via the detour paths in the pathways. 

Here, the path distance between input and output genes (i.e., the number of intermediate ML-genes between them) was irrelevant to the noise robustness. 
The shortest distance was similar for the three types, and the input and corresponding output genes were connected by one or two ML-genes. 
The noise was reduced not through a long-connected pathway but through multitrack, parallel paths.

\subsection*{Mutational robustness}

Robustness against the mutational changes in the GRN were then considered. Here, some paths, especially those that constituted a core structure, were inevitable for responses, but some other paths were not necessary to achieve high levels of fitness. The robustness against the removal of the latter was trivial. Thus, mutational robustness should be defined as a fitness change when indispensable paths were removed. Consequently, we studied mutational robustness against the number of removed paths that constituted a core structure (Fig~\ref{fig_mut} (c)). 

The Direct type had no mutational robustness, as the response was destroyed when any path that constituted a core structure was removed. The other two network types were more robust against mutational changes as their core structures contained detour paths. Even when a route connecting an input-output pair was disconnected, the external signal could still be transmitted by way of the detour paths. However, the FF-network and the Cooperative types were different in that the core structure of the former was separated according to each input-output pair, but not in the latter. Hence, their mechanisms of mutational robustness differed. In the FF-network type, when a path forming a core structure was removed, the response of a corresponding output gene dropped off, but the responses of the other output genes were hardly influenced. 
In contrast, in the Cooperative type with the LEGI structure, the responses of all output genes were weakened when any path in a core structure were removed. 
However, the decline was smaller as the Cooperative type included a larger number of alternative detour paths.

\begin{figure*}[htb]
\centering
\includegraphics[width=13.2 cm]{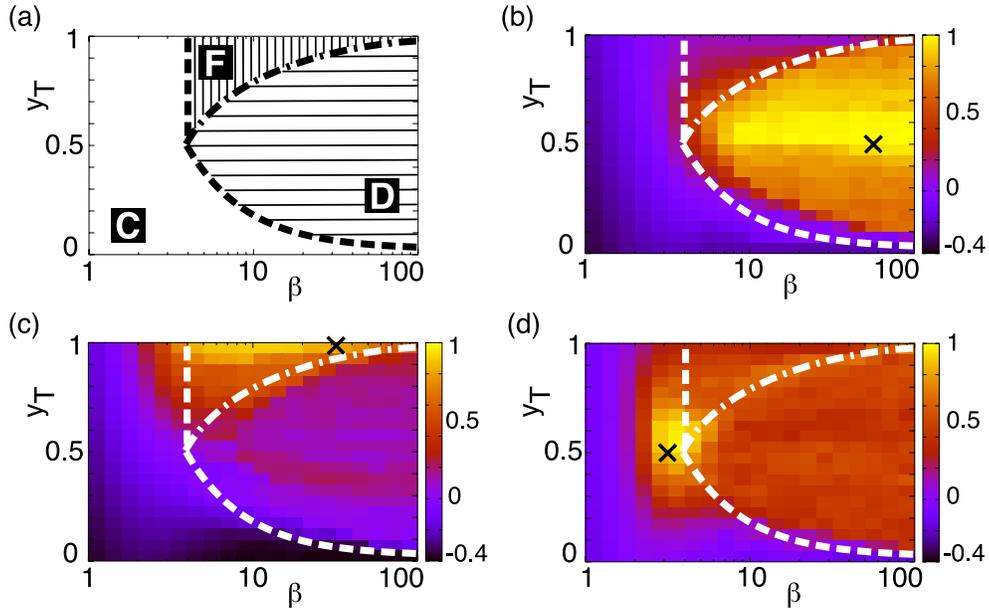}
\caption{{\bf Mutational robustness.}
(a) Phase diagram of the three types with regard to the $\beta$ (abscissa) and $y_T$ (ordinate) as estimated from the numerical simulations (the phase boundaries were obtained analytically~\cite{Inoue:2018aa}); Direct type (D: area with horizontal lines), FF-network type (F: area with vertical lines), and Cooperative type (C: blank area). 
(b--d) Changes in the fitness values against the shift in parameters. By using the network evolved from the parameters pointed by $\times$, fitness was computed for the parameter values ($\beta, y_T$) in the figure. (b) Direct type with $y_T = 0.5$ and $\beta = 10^{1.75}$, (c) FF-network type with $y_T = 1.0$ and $\beta = 10^{1.5}$, (d) Cooperative type with $y_T = 0.5$, and $\beta = \sqrt{10}$. The white curves indicate the phase boundaries.}
\label{fig_parameter}
\end{figure*}

In the evolutionary process so far, we have fixed each parameter (response sensitivity corresponding to the Hill coefficient $\beta$ and a constant threshold for expression $y_T$) of each gene to a common value and optimized the network structure for the given parameter values. 
We also studied the robustness to the changes in these parameter values (Fig~\ref{fig_parameter}). 
The Direct and FF-network types maintained a response only within a parameter range for each type, and the response often disappeared outside of their parameter range. On the other hand, the Cooperative type could maintain the response beyond the parameter range corresponding to its phase. 
This is because the LEGI structure includes the core structures for both the Direct and FF-network types within.

\subsection*{Tolerance to erroneous input} 

\begin{figure}[!h]
\centering
\includegraphics[width=7cm]{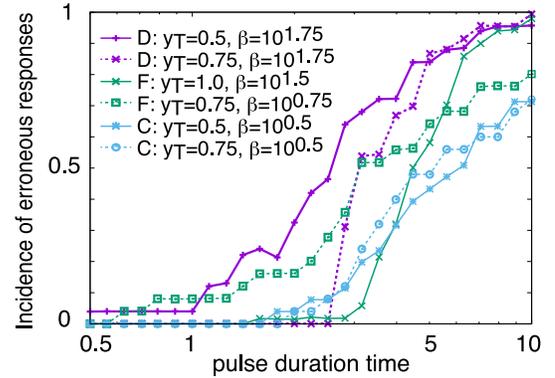}
\caption{{\bf Incidence of erroneous responses to a pulse input.}
The occurrence frequency of a response larger than 0.9 upon a pulse input by a corresponding target gene (ordinate) is plotted against the pulse duration time $\tau$ (abscissa). Results of the Direct type (purple $+$ and $\times$), the FF-network type (green $\times$ and $\square$), and the Cooperative type (blue $\ast$ and $\circ$), are shown using the same networks as in Fig~\ref{fig_mut}.}
\label{fig_Spulse}
\end{figure}

Thus far, the external input $S_{ext}$ was maintained at a steady value for $t\geq0$ in our model. Responses to a pulse input signal that was applied only for a brief time ($0\leq t\leq \tau$) were then investigated. 
These signals represent an erroneous input due to environmental fluctuations. It is of note that the time scale for the response dynamics of each gene were set to be scaled to $1$. 

The FF-network type was more tolerant than the Direct type with simpler motifs. Indeed, the feed-forwaed motif has been reported to be tolerant to pulse-type inputs~\cite{Alon:2006aa}. 
Interestingly, the Cooperative type was more tolerant to the pulse input than the FF-network type. In the Cooperative type, many ML-genes needed to change their expression levels in a sequential order before an output gene responded. An input gene stopped responding after a pulse interval, and the following ML-genes stopped responding before they stimulated the output gene. Thus, output genes avoided erroneous responses against a pulse input.

\subsection*{Adaptive responses to unforeseen challenges}

\begin{figure*}[htb]
\centering
\includegraphics[width=13.2 cm]{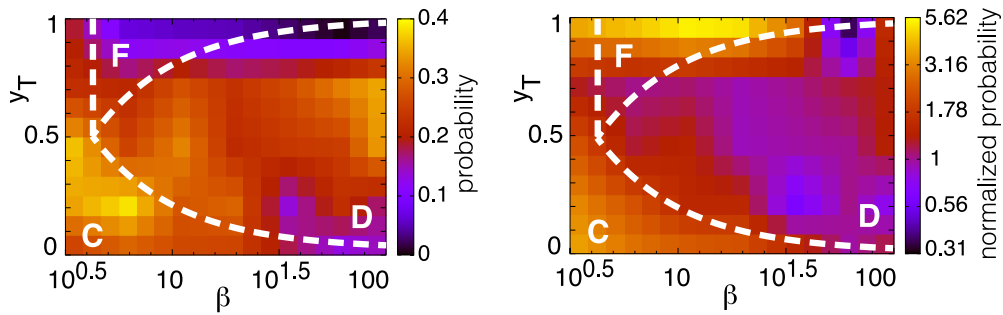} 
\caption{{\bf Responses to an unforeseen input.}
The response to an unforeseen input was calculated using the fittest networks evolved under each $\beta$ (abscissa) and $y_T$ (ordinate) values. The probability that the maximum response was larger than 0.01 (left) and the normalized probability compared with a case of random networks with each parameter value (right) are shown. The white curves show phase boundaries between the three types: Direct type (D), FF-network type (F), and Cooperative type (C).}
\label{fig_UCNR}
\end{figure*}

All living organisms must survive unforeseen environmental changes to avoid extinction. However, evolution is not an easy solution for survival, as it requires the expense of time. If there are no individuals adapted to a new environment prior to beneficial evolutionary changes, they will go extinct. 
It is essential for cells to respond to a new environment to some degree without evolutions rewiring of the GRN, and this is referred to as an `adaptive response to an unforeseen challenge (AUC)'~\cite{Braun:2015aa}. 

We studied the possibility of the AUCs using the following procedure: we prepared a GRN model by doubling the number of input and output genes but maintained the number of ML-genes. 
Using only half of the input-output pairs, networks were evolved to optimize their fitness, that is, to give postulated output responses to the inputs. 
Here, half of the input and output genes were not used for the fitness, and hence they were not related to the original evolution. After the original evolution, we studied the responses of unused output genes when one of the unused input genes was activated. The response for this non-evolutionary input without further changes in the GRN is shown in Fig~\ref{fig_UCNR}. 
Large responses to unforeseen inputs were observed in the Cooperative type networks, whereas there were almost no responses with the FF-network type. 
The Direct type had modest responses; however, this was because its core structure was so simple that even the addition of a random network could generate a response. 
In addition, such responses were rarer with larger network sizes. Hence, almost no response was expected in a network with many genes. 
The time course in the re-evolutionary process is shown in S1 Fig.

\subsection*{Resemblance of response dynamics between evolutionary and unforeseen inputs}

\begin{figure*}[htb]
\centering
\includegraphics[width=13.2 cm]{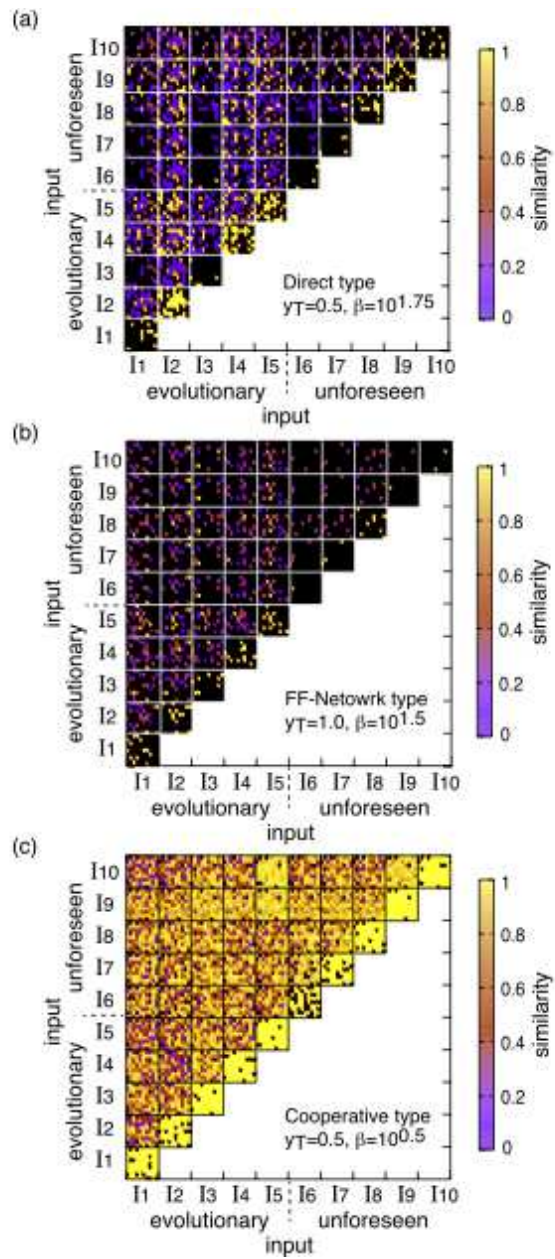}
\caption{{\bf Similarity in response dynamics against different inputs.}
Resemblance of the response dynamics of all ML-genes with each input signal calculated using dynamic time warping (DTW). Inputs $I_1 \sim I_5$ are original evolutionary conditions and $I_6 \sim I_{10}$ correspond to unforeseen inputs. Each square at the intersection shows the similarity of the ML-gene expression against $I_i$ and $I_j$. For each square, there are the number of ML-genes pixels and each pixel represents the distance between the ML-gene's responses upon $I_i$ and $I_j$ measured by DTW. Yellow indicates a complete correspondence (i.e., the responses of the gene upon $I_i$ and $I_j$ agree), purple indicates disagreement, and black indicates that when the gene shows no response to $I_i$ and $I_j$. (a) Direct type with $y_T = 0.5$ and $\beta = 10^{1.75}$, (b) FF-network type with $y_T = 1.0$ and $\beta = 10^{1.5}$, (c) Cooperative type with $y_T = 0.5$, and $\beta = \sqrt{10}$.}
\label{fig_dtw}
\end{figure*}

One reason why networks of the Cooperative type realize the AUC is that many ML-genes exhibit similar dynamics in response to the activation of any input gene.  
They also behave in a similar manner even when an unforeseen input gene that is not used for original evolution is activated. Here, we quantitatively studied the resemblance of the response dynamics of ML-genes with different input signals using dynamic time warping (DTW), as shown in Fig~\ref{fig_dtw}~\cite{Donald:1994aa,Eamonn:2001aa,Keogh:2005aa}. 
DTW analyzes the similarity between the data of the two time-series taking into account their elongation and parallel shifts in the time axis direction. 

For the Cooperative type, the ML-gene response dynamics to any input, including unforeseen ones, showed remarkable resemblance. These reflected cooperative behavior among the different inputs as well as among the different ML-genes. Although each input gene directly activated or inhibited a limited number of ML-genes as the network was sparse, almost all ML-genes subsequently responded through the entangled network with LEGI; some showed input specific responses, but most genes behaved independently of the input types. 
For the Direct and FF-network types, however, there were almost no such similarities among the different input signals that were observed, and only a small number of ML-genes exclusively responded to each input.

\subsection*{Dynamic mode decomposition of response dynamics} 

To investigate the origin of the resemblances of the response dynamics more closely, we analyzed the dynamics with dynamic mode decomposition (DMD)~\cite{Schmid:2010aa}. DMD decomposes time-series data with a large degree of freedom into a small number of dominant modes~\cite{Rowley:2009aa,Budisic:2012aa,Mezic:2013aa,Tu:2014aa,Brunton:2016aa}. It corresponds to a principle component analysis (PCA) for temporal sequence data. 

First, from a data series $\{ \mathbf{x}_0, \mathbf{x}_1, \cdots, \mathbf{x}_t \}$, we defined two time-series matrices $X=(\mathbf{x}_0, \mathbf{x}_1, \cdots, \mathbf{x}_{t-1})$ and $Y=(\mathbf{x}_1, \mathbf{x}_2, \cdots, \mathbf{x}_{t})$ and assumed a relationship of $Y=AX$ with an operator $A$. We calculated the eigenvalues and eigenvectors of $A$ by utilizing the singular value decomposition. When the dimension of each time sample $\mathbf{x}_i$ was $D_0$, $A$ was a square matrix of size $D_0$ and had $D_0$ pairs of eigenvalues and eigenvectors by definition. 
If only a small number ($r$) of singular values of $X$ had much larger values than others that were close to $0$, the singular value decomposition could be truncated to include only those $r$ modes. 
The $r$ modes dominantly influenced the dynamics of the data among the $D_0$ eigenvectors of $A$. These dominant eigenvectors corresponded to the primary axes in the PCA. 

\begin{figure*}[htb]
\centering
\includegraphics[width=13.2 cm]{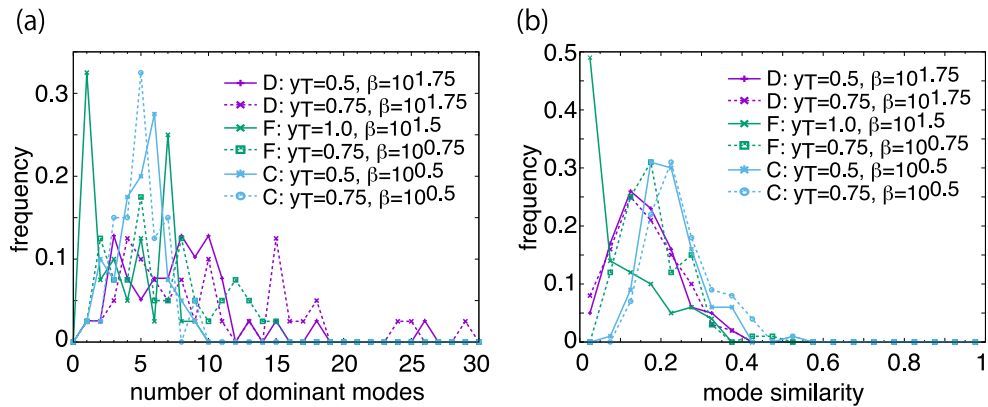} 
\caption{{\bf Response dynamics analyzed with dynamic mode decomposition.}
(a) Distribution of the number of dominant modes ($r$; abscissa). $r$ is defined as the number of singular values of a time-series matrix $X$ that are larger than 1. Four different fitted networks were used for each parameter set, and 10 time-series data sets were computed for each network corresponding to each input, including both evolutionary and unforeseen inputs. (b) Distribution of DMD mode similarity between modes for an evolutionary input and modes for an unforeseen input (abscissa). 
After computing 5 dominant DMD modes (eigenvectors) for each input, the mode similarity was obtained as the average value of the inner products between eigenvectors for respective inputs. For the time-series data corresponding to 5 evolutionary inputs and 5 unforeseen inputs, that is, $5 \times 5 =25$ combinations, the DMD mode similarity was computed to obtain the frequency histogram.}
\label{fig_dmd}
\end{figure*}

How the number of dominant modes ($r$) differed according to the network type was investigated first (see Fig~\ref{fig_dmd} (a)). For the Direct type, the number of dominant modes was almost the same as the number of responding ML-genes with each input. For the FF-network type, most ML-genes showed no responses to the unforeseen inputs, and hence $r=1$ truncation was just an artifact. In contrast, the Cooperative type showed responses that were represented by a smaller number of dominant modes. Its behavior could be described with fewer modes, although a larger number of ML-genes responded to each input. This result also indicated that ML-genes behaved cooperatively. It is of note that dimension reduction in high-dimensional adaptive states has been observed both in experiments and simulations~\cite{Kaneko:2015aa,Furusawa:2018aa}, whereas the results of this investigation provided a reduction in the response \textit{dynamics}. 

Finally, the similarity in the dominant modes depending on the different inputs was investigated. With truncated eigenvectors from time-series data for each input, an inner product between eigenvectors for an evolutionary input and for an unforeseen input was calculated and the distribution of the products was computed. This inner product took a larger (smaller) value when the eigenvectors were aligned (orthogonal), that is, when the dominant modes were similar (diverse). 
As shown in Fig~\ref{fig_dmd} (b), the Cooperative type showed the highest similarity, and the FF-network type had the lowest similarity. 
These results indicate that the response dynamics to an unforeseen input were like those that evolved with an evolutionary input for the Cooperative type. In other words, an unforeseen input was transmitted by utilizing an existing channel optimized for evolutionary inputs. 
Thus, the Cooperative type could respond to unforeseen inputs by activating any of the ML-genes in transfer of inputs.

\section*{Discussion}

In this paper, we have compared the three types of GRNs that exhibit appropriate input-output responses: Direct, FF-network, and Cooperative types. It was demonstrated that the last type with complex entangled networks showed cooperative responses involving many genes and were robust to noise, errors in the inputs, and network changes. 
Furthermore, it also showed adaptive responses to unforeseen challenges. Hence, this type of network has biological advantages over the others. 
Even though the other two simple unidirectional circuits, either direct or feed-forward networks, could show rapid and accurate responses, the entangled network with cooperative responses could better deal with the fluctuating and unforeseen environmental changes.

It is of note that complex networks with cooperative responses emerge when the expression responses of genes are sloppy, in other words, have a low Hill coefficient. 
This contrasts with the unidirectional simple networks that emerge when the gene responses are sensitive and accurate (with high Hill coefficient). 
In general, even after evolution, the Hill coefficients are not very high, and the expression of each gene is sloppy. 
The Cooperative type networks evolved to compensate for the sloppy and unreliable response of each gene. Accurate responses by an ensemble of unreliable sloppy elements was originally proposed by von Neumann for the design of computers. Here, as a byproduct, the present gene expression dynamics were robust to perturbations and showed adaptive responses to unforeseen challenges~\cite{Neumann:1956aa}.

The cellular networks are obviously entangled and composed of many elements that interact with each other. Cooperative responses have also been verified experimentally~\cite{Gasch:2000aa,Causton:2001aa,Stern:2007aa,Jozefczuk:2010aa,Gerashchenko:2012aa,Ho:2018aa}. Gasch et al. studied the diverse environmental stress responses of yeast by transcriptome analysis. 
They reported that certain sets of genes (approximately 900 genes) exhibited similar responses to almost all environmental changes, while some genes showed unique response patterns to specific conditions only~\cite{Gasch:2000aa}. 

One of the most remarkable properties of cells is their ability to adapt to unforeseen changes in environmental conditions ~\cite{Stolovicki:2006aa,David:2010aa,David:2013aa,Braun:2015aa}. 
Braun et al. found that yeast adapted to new environments that they had not previously experienced. 
This adaptive response occurred due to changes in their gene expression levels neither by means of prepared signaling networks nor by rewiring the networks via evolution. Despite the importance of such adaptive responses, how they are achieved remains elusive. 
The present investigation indicates a promising solution, by showing that the cooperative responses of the entangled networks induce similar responses for many genes for the experienced and unforeseen environmental inputs. 

As a constructive experiment, Isalan et al. explored the effect of adding genes, i.e., new links to the GRN of \textit{E.coli}~\cite{Isalan:2008aa,Baumstark:2015aa}. 
The new gene could provide unforeseen inputs (or perturbations) to the existing GRN, and could provide similar gene expression patterns and outputs as those that were accounted for by the existing GRNs. 
In addition, the gene expression patterns fell into a small number of attractors. 
This behavior is consistent with the observations for the Cooperative type, exhibiting similarity in their responses to unforeseen challenges with that to preexisting inputs. 

The origin of the correlated responses for many genes is due to local excitation specific to inputs and the global inhibition of many genes that are not specific to inputs. By considering the latter global response, the adaptive response to unforeseen challenges results. 
In biological systems, such local-excitation and global-inhibition (LEGI) is often observed, where inhibition occurs globally in space by using the global diffusion of inhibitors, for example, in chemotaxis pathways of eukaryotic cells~\cite{Takeda:2012aa} or \textit{Dictyostelium} cells~\cite{Wang:2012aa,Nakajima:2014aa}.  
In this investigation, global inhibition took place not in real space but in network space. We expect that this LEGI behavior can be detected experimentally by global analysis of gene expression patterns and cellular pathways. 

Cellular networks with many degrees of freedom are often studied by dividing them into small network motifs~\cite{Milo:2002aa,Shen-Orr:2002aa,Ma:2009aa}. It is then assumed that each motif has a characteristic function, whereas the function of the whole network is given as a summation of the functions of the involved motifs. However, paths in the network are entangled, and motifs that work independently are difficult to extract. Moreover, it is not clear whether motifs embedded into an entangled network have the same function as in isolation~\cite{Ma:2004aa,Jimenez:2017aa}.

At the same time, we do not necessarily state that the functions of network motifs are completely lost in an entangled network. Indeed, as already explained, the Cooperative type is robust against the changes in the parameter values because it includes the core structures of both the Direct and FF-network types (Fig~\ref{fig_parameter}). The Cooperative type can include such motifs that work in certain situations. 
In general studies, however, by taking advantage of the entangled network structure, the cooperative response is not decomposed into a combination of motifs.

In summary, the cooperative responses by entangled gene regulatory networks that are robust to noise and mutations can cope with unforeseen challenges, and this is essential for cell survival, as observed in recent microbial experiments.

\section*{Models}
The gene regulatory network model is composed of $N$ genes as nodes in a network, which are divided into three types: $N_{I}$ input genes receiving external inputs, $N_{O}$ output genes determining the fitness of the cell, and $N_{M}$ middle-layer genes (ML-genes) that transmit the input signals to the output genes ($N_{I} + N_{O} + N_{M} = N$). 
Through suitable normalization, the expression level of a gene is represented by a variable $x_i =[0,1]$ ($i=1, \dots, N$), with the maximal expression level scaled to unity.
The time evolution of each expression level is given as follows:  
\begin{equation}
   \frac{dx_i}{dt} = \frac{1}{1+\exp \left[-\beta (y_i - y_T)\right]} - x_i, \label{eq_dynamics} 
\end{equation} 
where $y_i = I_k \delta_{ik} + \sum_{j=1}^{N} C_{ij} x_j$ is the total signal received by the $i_{th}$ gene with $\delta_{ik}$ as the Kronecker delta for $k=1, \dots, N_I$. 
$I_k$ shows the external input to the $k_{th}$ input gene. 
$C_{ij}$ represents the regulation from gene $j$ to $i$, with $1$ (excitatory), $-1$ (inhibitory), and $0$ (non-existent). 
$y_T$ denotes a constant threshold, and $\beta$ determines the response sensitivity, which corresponds to the Hill coefficient. 
We assume that all genes in a network have the same $y_T$ and $\beta$ values for simplicity. 

For the evolution process, paths in the regulation matrix $C_{ij}$ were mutated, and $C_{ij}$ was selected according to the following fitness condition: for each given input, one predominant target among the output genes was expressed. Specifically, only the gene $N-N_O+k$ among the output genes $N-N_O+1, \cdots, N$ should respond upon the application of input $I_k$ ($k=1, \dots, N_I$), where the response was given as the difference between the final ($\overline{x_{i}^{fin}}$) and initial ($\overline{x_{i}^{ini}}$) expression levels averaged over a time span, respectively, as 
\begin{align}
   fitness = \frac{1}{N_I} \sum_{k=1}^{N_I} &(\overline{x_{N-N_O+k}^{fin}}- \overline{x_{N-N_O+k}^{ini}}  \notag \\ 
                  & - \frac{1}{N_O -1} \sum_{j=1, j \ne k}^{N_O} \overline{x_{N-N_O+j}^{fin}} ).  
 \label{eq_fitness}
\end{align}
With this fitness function, we evolved the network structure, \textit{that is} the regulation matrix $C_{ij}$ by using a simple genetic algorithm. 
In the mutation process, we fixed the number of paths and swapped the connection $C_{ij}$ with a small mutation rate.

\section*{Supporting information}

\paragraph*{S1 Fig.}
\label{S1_Fig}
{\bf Time course in re-evolutionary process.} 
Changes in fitness (ordinate) over a re-evolutionary course with an unforeseen input against the generation (abscissa). The networks are evolved to select a larger response to the new target gene corresponding to the unforeseen input. The averaged fitness from 125 different trials are plotted for each parameter set. Networks are re-evolved keeping the number of total paths from the fitted structure after the original evolution (a) or from its core structure (b).

\section*{Acknowledgments}
The authors would like to thank Macoto Kikuchi, Namiko Mitarai, Naoki Honda, Chikara Furusawa, Nobuhiko J. Suematsu, and Hirokazu Ninomiya for their stimulating discussion. This research was partially supported by a Grant-in-Aid for Scientific Research (A) (20H00123) and  Grant-in-Aid for Scientific Research on Innovative Areas (17H06386) from the Ministry of Education, Culture, Sports, Science and Technology (MEXT) of Japan.

\bibliography{Papers-plos}

\end{document}